\documentclass[12pt]{article}

\begin{document}

\baselineskip 20pt
\vspace*{-.6in}
\thispagestyle{empty}
\begin{flushright}
CALT-68-2335\\ CITUSC/01-025
\end{flushright}

\vspace{.5in} {\Large
\begin{center}
{\bf Anomaly Cancellation: A Retrospective From a Modern Perspective}\end{center}}
\bigskip
\begin{center}
 John H. Schwarz\\
\bigskip
\emph{California Institute of Technology, Pasadena, CA  91125, USA\\
and\\
Caltech-USC Center for Theoretical Physics\\
University of Southern California, Los Angeles, CA 90089, USA}
\end{center}
\vspace{0.8in}

\begin{center}
\textbf{Abstract}
\end{center}
\begin{quotation}
\noindent The mechanism by which gauge and gravitational
anomalies cancel in certain string theories is reviewed.  The
presentation is aimed at theorists who do not necessarily
specialize in string theory.
\end{quotation}

\vspace{0.8in}

\centerline{Talk presented at {\it 2001: A Spacetime Odyssey} -- the inaugural conference}
\centerline{of the Michigan Center for Theoretical Physics}

\newpage

\pagenumbering{arabic}
\section{Introduction}

As is well-known, in the early 1980s it appeared that superstrings
could not describe parity-violating theories, because of quantum
mechanical inconsistencies due to anomalies. The discovery that in
certain cases the anomalies could cancel\cite{Green:1984sg} was
important for convincing many theorists that string theory is a
promising approach to unification. In the 17 years that have
passed since then, string theory has been studied intensively, and
many issues are understood much better now. This progress enables
me to describe the original anomaly cancellation mechanism in a
more elegant way than was originally possible. The
improvements that are incorporated in the following discussion
include an improved understanding of the association of specific
terms with specific string world-sheets as well as improved tricks
for manipulating the relevant characteristic classes.

When a symmetry of a classical theory is broken by radiative corrections,
the symmetry is called ``anomalous.'' In this case there is no
choice of local counterterms that can be added to the low energy
effective action to restore the symmetry. Anomalies arise from
divergent Feynman diagrams, with a classically conserved current
attached, that do not admit a regulator compatible with
conservation of the current. Anomalies only arise at one-loop
order (Adler-Bardeen theorem) in diagrams with a chiral fermion or
boson going around the loop. Their origin can be traced (Fujikawa)
to the behavior of Jacobian factors in the path-integral measure.

There are two categories of anomalies. The first category consists
of anomalies that break a global symmetry. An example is the axial
part of the flavor $SU(2) \times SU(2)$ symmetry of QCD.  These
anomalies are ``good'' in that they do not imply any
inconsistency. Rather, they make it possible to carry out certain
calculations to high precision. The classic example is the rate for
the decay $\pi^0 \to \gamma\gamma$. The second category of
anomalies consists of ones that break a local gauge symmetry.
These are ``bad'', in that they imply that the quantum theory is
inconsistent. They destroy unitarity, causality, and other related
sacred principles. In the remainder of this talk, I will only be
concerned with this second category of anomalies. Either they
cancel, or the theory in question is inconsistent.

Chiral fields only exist in spacetimes with an even dimension. If
the dimension is $D= 2n$, then anomalies can occur in diagrams
with one current and $n$ gauge fields attached to a chiral field
circulating around the loop. In four dimensions these are triangle
diagrams and in ten dimensions these are hexagon diagrams. The
resulting nonconservation of the current $J^{\mu}$ takes the form
\begin{equation}
\partial_{\mu} J^{\mu} \sim \epsilon^{\mu_1 \mu_2 \ldots \mu_{2n}}
F_{\mu_1 \mu_2}  \cdots
F_{\mu_{2n-1} \mu_{2n}}
\end{equation}

In string theory there are various world-sheet topologies that
correspond to one-loop diagrams. In the case of  type II or
heterotic theories it is a torus. For the type I superstring
theory it can be a torus, a Klein bottle, a cylinder or a M\"obius
strip. However, the anomaly analysis can be carried out entirely
in terms of a low-energy effective field theory, which is what I
will do. Still it is interesting to interpret the Type I result in
terms of string world sheets. The torus turns out not to
contribute to the anomaly. For the other world sheet topologies,
it is convenient to imagine them as made by piecing together
boundary states $|B\rangle$ and cross-cap states $|C\rangle$. (Cross-caps can
be regarded as boundaries that have opposite points identified.)
In this way $\langle B|B\rangle$ represents a cylinder, 
$\langle B|C\rangle$ and $\langle C|B \rangle$
represent a M\"obius strip, and $\langle C|C\rangle$ represents a Klein
bottle. The correct relative weights are encoded in the
combinations
\begin{equation}
(\langle B| + \langle C|) \times (|B\rangle  + |C\rangle).
\end{equation}
We will interpret the consistency of the $SO(32)$ type I theory as
arising from a cancellation between the boundary and cross-cap
contributions. It should also be pointed out that the modern
interpretation of the boundary state is in terms of a world-sheet
that ends on a {\it D-brane}, whereas the cross-cap state
corresponds to a world-sheet that ends on an {\it orientifold
plane}.

\section{Anomaly Analysis}
\subsection{Chiral Fields}
The kinds of chiral fields that can exist depend on the number of
space and time dimensions with a pattern that repeats every 8
dimensions. For example, it is true in both four and ten
dimensions that a fermion field can be Majorana (real in a
suitable representation of the Dirac algebra) or Weyl (definite
handedness). However, a significant difference between the two
cases is that, unlike the case of four dimensions,
in ten dimensions the two conditions are compatible,
so that it is possible for a fermion field to be simultaneously
Majorana and Weyl. Indeed, the basic (irreducible) spinors in ten
dimensions are Majorana--Weyl. These statements depend on the
signature as well as the dimension. I am assuming Lorentzian
signature throughout.

Another difference between four and ten dimensions is that in ten
dimensions it is also possible to have chiral bosons! To be
specific, consider a fourth rank antisymmetric tensor field
$A_{\mu\nu\rho\lambda}$ , which is conveniently represented as a
four-form A. Then the five-form field strength $F=dA$ has a gauge
invariance analogous to that of the Maxwell field.  Moreover, one
can covariantly eliminate half of the degrees of freedom
associated with this field by requiring that it is self-dual (or
anti-self dual). The resulting degrees of freedom are not
reflection invariant, and they therefore describe a chiral boson.
The self-duality condition of the free theory is deformed by interaction terms.
This construction in ten dimensions is consistent with Lorentzian
signature, whereas in four dimensions a two-form field strength
can be self-dual for Euclidean signature (instantons).

\subsection{Differential Forms and Characteristic Classes}
To analyze anomalies it is extremely useful to use differential
forms and characteristic classes. In modern times this kind of
mathematics has become part of the standard arsenal of theoretical
physicists.  For example, Yang--Mills fields are
Lie-algebra-valued one-forms:
\begin{equation}
A = \sum_{\mu, a} A^{a}_{\mu}(x) \lambda^{a} dx^{\mu}.
\end{equation}
Here the $\lambda^a$ are matrices in a convenient representation
(call it $\rho$) of the Lie algebra ${\cal G}$. The field
strengths are Lie-algebra-valued two-forms:
\begin{equation}
F = {1\over2} \sum_{\mu\nu} F_{\mu\nu} dx^{\mu} \wedge dx^{\nu} = dA + A\wedge A.
\end{equation}
Under an infinitesimal gauge transformation
\begin{equation}
\delta_{\Lambda} A = d\Lambda + [A, \Lambda],
\end{equation}
\begin{equation}
\delta_{\Lambda} F = [F, \Lambda].
\end{equation}
$\Lambda$ is an infinitesimal Lie-algebra-valued zero-form.

Gravity (in the vielbein formalism) is described in an almost identical manner. The
{\it spin connection} one-form
\begin{equation}
\omega = \sum_{\mu, a} \omega^{a}_{\mu}(x) \lambda^{a} dx^{\mu}.
\end{equation}
is a gauge field for local Lorentz symmetry. The $\lambda^a$ are chosen to be in
the fundamental representation of the Lorentz algebra ($D\times D$ matrices).
The curvature two-form is
\begin{equation}
R =  d\omega + \omega\wedge \omega.
\end{equation}
Under an infinitesimal local Lorentz transformation (with
infinitesimal parameter $\Theta$)
\begin{equation}
\delta_{\Lambda} \omega = d\Theta + [\omega, \Theta],
\end{equation}
\begin{equation}
\delta_{\Lambda} R = [R, \Theta].
\end{equation}

Characteristic classes are differential forms, constructed out of
$F$ and $R$, that are closed and gauge invariant. Thus $X(F,R)$ is
a characteristic class provided that $dX=0$ and $\delta_{\Lambda}
X = \delta _{\Theta} X =0$. Some examples are
\begin{equation}
{\rm tr}(F\wedge \ldots \wedge F) \equiv {\rm tr}(F^k),
\end{equation}
\begin{equation}
{\rm tr}(R\wedge \ldots \wedge R) \equiv {\rm tr}(R^k),
\end{equation}
as well as polynomials built out of these building blocks using wedge products.

\subsection{Characterization of Anomalies}
Yang--Mills and local Lorentz anomalies in $D=2n$ dimensions are
encoded in a characteristic class that is a $2n+2$ form, denoted
$I_{2n+2}$.  You can't really antisymmetrize $2n+2$ indices in
$2n$ dimensions, so these expressions are a bit formal, though
they can be given a precise mathematical justification. In any
case, the physical anomaly is characterized by a $2n$ form $G_{2n}$,
which certainly does exist. The precise formula is
\begin{equation}
\delta S_{\rm eff} = \int G_{2n}.
\end{equation}
The formulas for $G_{2n}$ are rather ugly and subject to the
ambiguity of local counterterms and total derivatives, 
whereas by pretending that there
are two extra dimensions one uniquely encodes the anomalies in
beautiful formulas $I_{2n+2}$. Moreover, any $G_{2n}$ that is
deduced from an $I_{2n+2}$ by the formulas that follow,  is
guaranteed to satisfy the Wess--Zumino consistency conditions.

The anomaly $G_{2n}$ is obtained from $I_{2n+2}$ (in a coordinate
patch) by the descent equations $I_{2n+2} = d\omega_{2n+1}$ and
$\delta \omega_{2n+1} = d G_{2n}$. Here $\delta$ represents a
combined gauge transformation (i.e., $\delta = \delta_{\Lambda} +
\delta_{\Theta}$). The ambiguities in the determination of the
Chern--Simons form $\omega_{2n+1}$ and the anomaly form $G_{2n}$
from these equations are just as they should be and do not pose a
problem. The total anomaly is a sum of contributions from each of
the chiral fields in the theory, and it can be encoded in a
characteristic class
\begin{equation}
I_{2n+2} = \sum_{\alpha} I^{(\alpha)}_{2n+2}
\end{equation}

The formulas for every possible anomaly contribution
$I^{(\alpha)}_{2n+2}$ were worked out by Alvarez-Gaum\'e and
Witten.\cite{Alvarez-Gaume:1984ig} Dropping an overall
normalization factor (which we can do, because we are interested
in achieving cancellation) their results are as follows:

\noindent $\bullet$ A left-handed Weyl fermion belonging to the
$\rho$ representation of the Yang-Mills gauge group contributes
\begin{equation}
I_{1/2}(R,F) = \left( \hat A (R)\,  {\rm tr}_{\rho} e^{iF} 
\right)_{2n+2}    
\, .
\end{equation}
The notation $(\cdots)_{2n+2}$ means that one should extract the
$(2n+2)$-form part of the enclosed expression. The Dirac roof
genus $\hat A(R)$ is given by
\begin{equation} \label{Aeqn}
\hat A(R) = \prod_{i=1}^n \left({\lambda_i /2 \over {\rm sinh}
\lambda_i /2} \right),
\end{equation}
where the $\lambda_i$ are the ``eigenvalue two-forms'' of the
curvature:
\begin{equation}
R \sim  \pmatrix{ 0&\lambda_1& & & & &&\cr
-\lambda_1&0& & & & &&\cr
 & &0&\lambda_2 & & &&\cr
 & &-\lambda_2&0 & & &&\cr
 & & & & . & &&\cr
& & & & &. &&\cr
& & & & && 0 & \lambda_n\cr
& & & & & & -\lambda_n & 0\cr} .
\end{equation}

\noindent $\bullet$ A left-handed Weyl gravitino, which is always
a singlet of any Yang-Mills groups, gives a contribution denoted
$I_{3/2}(R)$. In the following, we will circumvent the need for
the explicit formula.

\noindent $\bullet$ A self-dual tensor gives a contribution
denoted $I_A (R)$. It is related to the Hirzebruch $L$-function
\begin{equation} \label{Leqn}
L(R) = \prod_{i=1}^n {\lambda_i \over {\rm tanh} \lambda_i}
\end{equation}
by $I_A (R) = -{1 \over 8} L(R)$.

In each case a chiral field of the opposite chirality
(right-handed instead of left-handed) gives an anomaly
contribution of the opposite sign. Later we will utilize the
identity\cite{Scrucca:1999jq}
\begin{equation} \label{magiceq}
\hat A(R/2) = \sqrt{L(R/4) \hat A (R)},
\end{equation}
which is an immediate consequence of eqs. (\ref{Aeqn}) and
(\ref{Leqn}).

\subsection{The Type IIB Theory}

Type IIB superstring theory is a ten-dimensional parity-violating
theory, whose massless chiral fields consist of two left-handed
Majorana--Weyl gravitinos (or, equivalently, one Weyl gravitino),
two right-handed Majorana--Weyl spinors (or ``dilatinos'') and a
self-dual boson. Thus the total anomaly is given by the 12-form
part of
\begin{equation}
I = I_{3/2}(R) - I_{1/2}(R) + I_A (R).
\end{equation}
An important result of the Alvarez-Gaum\'e and Witten paper
\cite{Alvarez-Gaume:1984ig} is that this 12-form vanishes, so that
this theory is anomaly-free. The proof requires showing that the
expression
\begin{equation}
\left( \prod_{i=1}^5 {\lambda_i/2 \over {\rm sinh} \lambda_i /2}
\right) \left( 2\sum {\rm cosh} \lambda_i -2 \right) - {1 \over 8}
\prod_{i=1}^5 {\lambda_i \over {\rm tanh} \lambda_i}
\end{equation}
contains no terms of sixth order in the $\lambda_i$. This involves
three nontrivial cancellations. The relevance of this fact to the
type I theory, to which we turn next, is that we can represent
$I_{3/2} (R)$ by $I_{1/2}(R) - I_A (R)$. This is only correct for
the 12-form part, but that is all that we need.

\section{Type I Superstring Theory}
Type I superstring theory has 16 conserved supercharges, which
form a Majorana--Weyl spinor in ten dimensions. The massless
fields of type I superstring theory consist of a supergravity
multiplet in the closed string sector and a super Yang--Mills
multiplet in the open string sector. The supergravity multiplet
contains three bosonic fields: the metric (35), a two-form (28),
and a scalar dilaton (1). The parenthetical numbers are the number
of physical polarization states represented by these fields. None
of these is chiral. It also contains two fermionic fields: a
left-handed Majorana--Weyl gravitino (56) and a right-handed
Majorana--Weyl dilatino (8). These are chiral and contribute an
anomaly given by
\begin{equation}
I_{\rm sugra} = {1 \over 2} \left( I_{3/2} (R) - I_{1/2}
(R)\right)_{12} = - {1\over 2} \left( I_A(R) \right)_{12} = {1
\over 16} \left( L(R) \right)_{12}.
\end{equation}

The super Yang--Mills multiplet contains the gauge fields and
left-handed Majorana--Weyl fermions (gauginos), each of which belongs to the
adjoint representation of the gauge group. Classically, the gauge
group can be any orthogonal or symplectic group. In the following
we only consider the case of $SO(N)$, since it is the one of most
interest. In this case the adjoint representation corresponds to
antisymmetric $N \times N$ matrices, and has dimension $N(N-1)/2$.
Adding the anomaly contribution of the gauginos to the
supergravity contribution given above yields
\begin{equation} \label{typeIanomaly}
I_{12} = \left( {1\over2} \hat A (R)\, {\rm Tr} e^{iF} + {1 \over
16} L(R) \right)_{12}
\end{equation}
The symbol Tr is used to refer to the adjoint representation,
whereas the symbol tr is used (later) to refer to the
$N$-dimensional fundamental representation.

Next we use the Chern character property
\begin{equation}
{\rm tr}_{\rho_1 \times \rho_2} e^{iF} = \left({\rm tr}_{\rho_1 }
e^{iF} \right) \left( {\rm tr}_{\rho_2} e^{iF} \right)
\end{equation}
to deduce that for $SO(N)$
\begin{equation} \label{Chern}
{\rm Tr} e^{iF} = {1 \over 2} \left( {\rm tr} e^{iF} \right)^2 -
{1 \over 2} {\rm tr} e^{2iF} = {1 \over 2} \left( {\rm tr} \, {\rm
cos}  F \right)^2 - {1 \over 2} {\rm tr}\, {\rm cos} 2F.
\end{equation}
In the last step we have used the fact that the trace of an odd
power of $F$ vanishes.

Substituting eq. (\ref{Chern}) into eq. (\ref{typeIanomaly}) gives
the anomaly as the 12-form part of
\begin{equation}
 {1 \over 4}\hat A (R) \left( {\rm tr} \, {\rm
cos}  F \right)^2 - {1 \over 4} \hat A(R) {\rm tr}\, {\rm cos} 2F
+ {1 \over 16} L(R) .
\end{equation}
Since this is of 6th order in $R$'s and $F$'s, the following
expression has the same 12-form part:
\begin{equation}
I' = {1 \over 4}\hat A (R) \left( {\rm tr} \, {\rm cos}  F
\right)^2 - 16 \hat A(R/2) {\rm tr}\, {\rm cos} F + 256 L(R/4) .
\end{equation}
Moreover, using eq. (\ref{magiceq}), this can be recast as a
perfect square
\begin{equation}
I' = \left( {1 \over 2}\sqrt{\hat A (R)}  {\rm tr} \, {\rm cos} F
- 16 \sqrt{ L(R/4)} \right)^2 .
\end{equation}

There is no choice of $N$ for which $I_{12}^{\prime} = I_{12}$
vanishes. However, as will be explained later, it is possible to
introduce a local counterterm that cancels the anomaly if $I_{12}$
factorizes into a product of a 4-form and an 8-form. We have shown
that $I' = Y^2$, where
\begin{equation}
Y =  {1 \over 2}\sqrt{\hat A (R)} \, {\rm tr} \, {\rm cos} F - 16
\sqrt{ L(R/4)} .
\end{equation}
A priori, this is a sum of forms $Y_0 + Y_4 + Y_8 + \ldots$.
However, if the constant term vanishes ($Y_0 = 0$), then
\begin{equation}
I_{12} = \left(Y_4 + Y_8 + \ldots \right)_{12}^2 = 2 Y_4 Y_8
\end{equation}
as required. To examine the constant term in $Y$, we use the fact
that $L$ and $\hat A$ are equal to 1 plus higher order forms and
that ${\rm tr} \, {\rm cos}\, F = N + \ldots$ to deduce that $Y_0
= (N - 32 )/2$. Thus, the desired factorization only works for the
choice $N = 32$ in which case the gauge algebra is $SO(32)$.

Let us express $Y$ as a sum of two terms $Y_B + Y_C$, where
\begin{equation}
Y_B =  {1 \over 2}\sqrt{\hat A (R)} \, {\rm tr} \, {\rm cos} F
\end{equation}
and
\begin{equation}
Y_C =  - 16 \sqrt{ L(R/4)} .
\end{equation}
This decomposition has a simple interpretation in terms of string
world-sheets. $Y_B$ is the boundary -- or D-brane -- contribution.
It carries all the dependence on the gauge fields. $Y_C$ is the
cross-cap -- or orientifold plane -- contribution. Note that
\begin{equation}
I' = Y^2 = Y_B^2 + 2Y_B Y_C + Y_C^2
\end{equation}
displays the anomaly contributions arising from distinct
world-sheet topologies: the cylinder, the M\"obius strip, and the
Klein bottle.

In order to cancel the anomaly, what we need is a local
counterterm, $S_{C}$, with the property that
\begin{equation} \label{varycounter}
\delta S_{C} = - \int G_{10},
\end{equation}
where $G_{10}$ is the anomaly  10-form that follows, via the
descent equations, from $I_{12} = 2 Y_4 Y_8$. As we mentioned
earlier, there are ambiguities in the determination of $G_{10}$
from $I_{12}$. A convenient choice in the present case is 
\begin{equation}
G_{10} =  2 G_2 Y_8,
\end{equation} 
where $G_2$ is a two-form that is related to $Y_4$ by
the descent equations $Y_4 = d \omega_3$ and $\delta \omega_3 = d G_2$. 
This works because $Y_8$ is closed and
gauge invariant.

Recall that the type I supergravity multiplet contains a two form
field, which we denote $C_2$. (Parenthetically, we note that in
the lingo of the RNS superstring description it belongs to the
Ramond-Ramond sector of the spectrum.) In terms of its index
structure, it would seem that the field $C$ should be invariant
under Yang--Mills gauge transformations and local Lorentz
transformations. However, it does transform nontrivially under
each of them in just such a way as to cancel the
anomaly.\cite{Green:1984sg} Specifically, writing the counterterm
as
\begin{equation}
 S_{C} = \mu \int C_2 Y_8,
\end{equation}
eq. (\ref{varycounter}) is satisfied provided that
\begin{equation}
\mu \delta C_2 = -2 G_2.
\end{equation}
The coefficient $\mu$ is a parameter whose value depends on
normalization conventions that we are not specifying here.

One consequence of the nontrivial gauge transformation properties
of the field $C_2$ is that the naive kinetic term $ \int |dC_2|^2$
must be modified to give gauge invariance. The correct choice is
$\int |H_3 |^2$, where
\begin{equation}
H_3 = dC_2 + 2 \mu^{-1} \omega_3.
\end{equation}
Note that $\omega_3$ contains both  Yang--Mills and Lorentz
Chern--Simons terms. Only the former is present in the classical
supergravity theory.

\section{Concluding Remarks}
The techniques that I have described for analyzing and cancelling
anomalies in the type I $SO(32)$ theory can be used to
analyze more complicated examples. Recently, in work carried out
with Witten,\cite{Schwarz:2001sf} a number of other examples were
analyzed. These included the following

\noindent $\bullet$ The type IIB theory with $n$ spacetime-filling
D-brane anti-D-brane pairs and gauge symmetry $U(n) \times
U(n)$.\cite{Srednicki:1998mq} This model has tachyon fields
associated to open strings connecting D-branes to anti-D-branes.

\noindent $\bullet$ The type I theory with $n$ additional
spacetime-filling D-brane anti-D-brane pairs and gauge symmetry
$SO(32+n) \times SO(n)$.\cite{Sugimoto:1999tx}

\noindent $\bullet$ A tachyon-free ten-dimensional string model
(originally due to Sugimoto\cite{Sugimoto:1999tx}) with
spontaneously broken supersymmetry and gauge group $Sp(16)$.

\noindent $\bullet$ A nonsupersymmetric and tachyon-free
ten-dimensional string model (originally due to
Sagnotti\cite{Sagnotti:1995ga}) with gauge group $U(32)$.

\noindent $\bullet$ Various six-dimensional string models.

\section*{Acknowledgments}
This work was supported in part by the U.S. Dept. of Energy under
Grant No. DE-FG03-92-ER40701.

\end{document}